\documentclass[twoside,journey]{IEEEtran}
\usepackage{makecell}
\usepackage{hyperref}
\usepackage{array}
\usepackage{graphicx,amssymb,amsmath}
\usepackage{multicol}
\usepackage[noadjust]{cite}
\usepackage{setspace}
\usepackage{subfigure}
\usepackage{graphicx}
\usepackage{float}
\usepackage{url}
\usepackage{stfloats}
\usepackage{amsthm,pifont}
\usepackage{flushend}
\usepackage{cases,subeqnarray}
\usepackage{bm,multirow,bigstrut}
\usepackage{amsmath, amsthm, amssymb}
\usepackage{textcomp}
\usepackage{latexsym,bm}
\usepackage{booktabs}
\usepackage{xcolor}
\usepackage{mathtools}
\usepackage{dsfont}
\usepackage{extarrows}
\usepackage{epsfig}
\usepackage{epsfig}
\usepackage{epstopdf}
\usepackage[noend]{algpseudocode}
\usepackage{algorithmicx,algorithm}
\theoremstyle{plain}

\theoremstyle{plain}
\newtheorem{rem}{Remark}

\usepackage{amsmath}

\IEEEoverridecommandlockouts

\begin{document}
\title{Rethinking Wireless Communication Security in Semantic Internet of Things}
\author{Hongyang~Du, Jiacheng~Wang, Dusit~Niyato,~\IEEEmembership{Fellow,~IEEE}, Jiawen~Kang, Zehui~Xiong, Mohsen~Guizani,~\IEEEmembership{Fellow,~IEEE}, and Dong~In~Kim,~\IEEEmembership{Fellow,~IEEE}
\thanks{H.~Du, J.~Wang and D.~Niyato are with the School of Computer Science and Engineering, Nanyang Technological University, Singapore (e-mail: hongyang001@e.ntu.edu.sg, jcwang\_cq@foxmail.com, dniyato@ntu.edu.sg).}
\thanks{J. Kang is with the School of Automation, Guangdong University of Technology, China. (e-mail: kavinkang@gdut.edu.cn).}
\thanks{Z. Xiong is with the Pillar of Information Systems Technology and Design, Singapore University of Technology and Design, Singapore (e-mail: zehui\_xiong@sutd.edu.sg).}
\thanks{M. Guizani is with Machine Learning Department, Mohamed Bin Zayed University of Artificial Intelligence, Abu Dhabi, UAE (e-mail: mguizani@ieee.org).}
\thanks{D. I. Kim is with the Department of Electrical and Computer Engineering, Sungkyunkwan University, South Korea (e-mail: dikim@skku.ac.kr).}
}
\maketitle
\vspace{-1cm}
\begin{abstract}
Semantic communication is an important participant in the next generation of wireless communications. Enabled by this novel paradigm, the conventional Internet-of-Things (IoT) is evolving toward the semantic IoT (SIoT) to achieve significant system performance improvements. However, traditional wireless communication security techniques for bit transmission cannot be applied directly to the SIoT that focuses on semantic information transmission. One key reason is the lack of new security performance indicators. Thus, we have to rethink the wireless communication security in the SIoT. As such, in the paper, we analyze and compare classical security techniques, i.e., physical layer security, covert communications, and encryption, from the perspective of semantic information security. We highlight the differences among these security techniques when applied to the SIoT. Novel performance indicators such as semantic secrecy outage probability (for physical layer security techniques) and detection failure probability (for covert communication techniques) are proposed. Considering that semantic communications can raise new security issues, we then review attack and defense methods at the semantic level. Finally, we present several promising directions for future secure SIoT research.
\end{abstract}
\IEEEpeerreviewmaketitle
\section{Introduction}
\IEEEPARstart{T}{he} advancement of semantic communications technique has brought about significant changes to nearly all aspects of wireless communication networks. As a new communication paradigm, semantic communications no longer focus on the accurate transmission of bits, but on the transmission of task-related semantic information~\cite{xu2022edge}. As semantic models become lightweight, deploying semantic encoders and decoders in network edge devices is also practical. Thus, the conventional Internet-of-Things (IoT) is evolving toward the semantic IoT (SIoT), enabling more efficient and energy-saving information interaction~\cite{yang2021energy}.

\begin{figure*}[t]
\centering
\includegraphics[width = 0.98\textwidth]{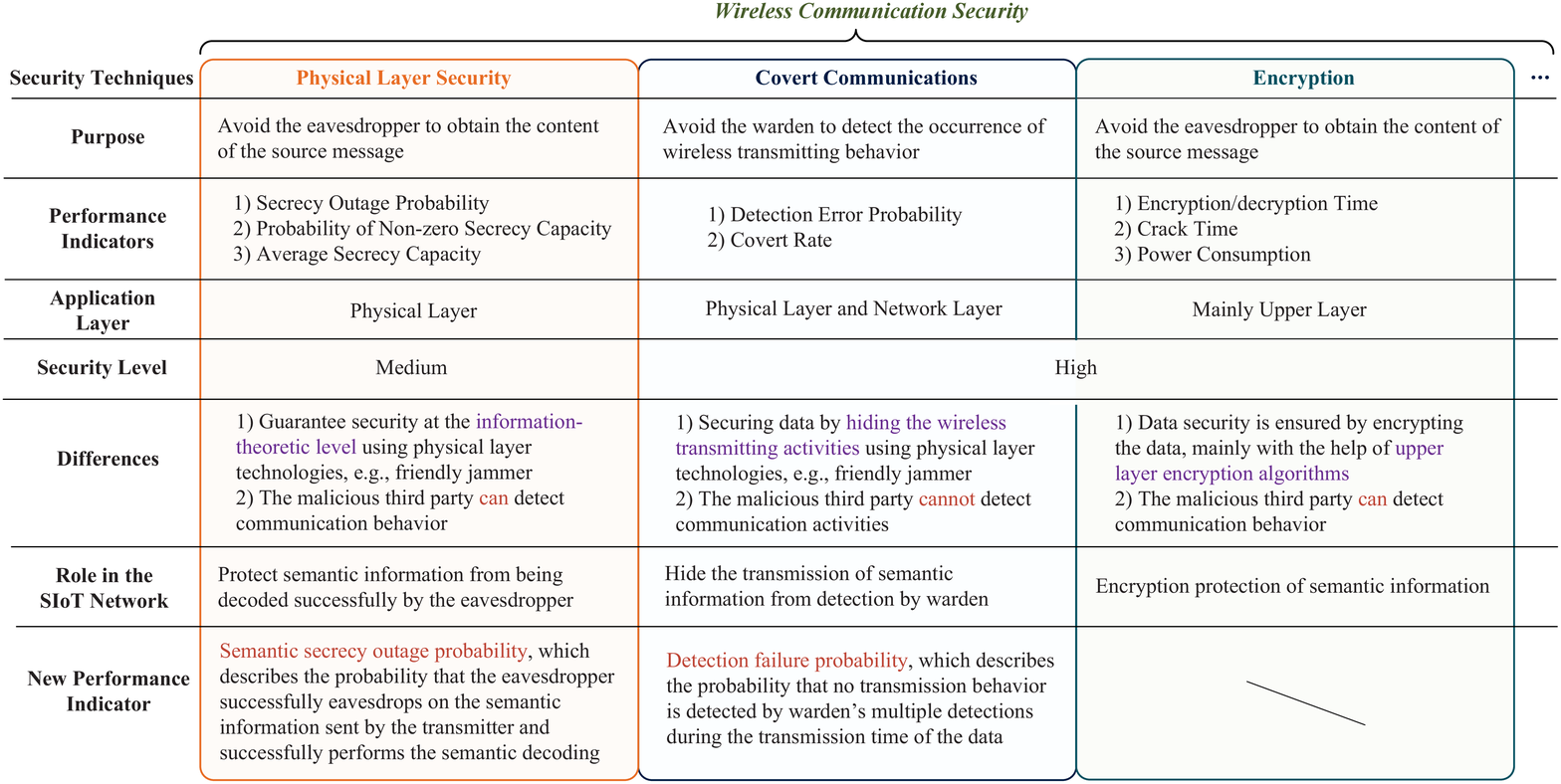}
\caption{Comparison of physical layer security, covert communications, and encryption techniques, and our proposed new performance indicators for the SIoT.}
\label{Visio-Compare}
\end{figure*}

Data security is an enduring topic in wireless communication networks. When network devices transmit sensitive data, there is a high risk of eavesdropping or jamming/interference attacks by malicious third parties. A straightforward approach is to encrypt the data by using complex algorithms. Although various encryption algorithms have been proposed, the effectiveness of encryption and the complexity of computation are  positively correlated. To improve the security of communication networks, high complexity in encryption put much pressure on network edge devices with insufficient computing power. Fortunately, besides traditional encryption methods, the physical layer security (PLS) technique is considered to be an essential driver enhancing 6G security~\cite{chorti2022context}. Without requiring actual key distribution, PLS can achieve high-quality network security performance with low computational complexity. A common disadvantage of encryption and PLS is that they can only guarantee that the information will not be decrypted. However, a malicious third party can still use the detected wireless signals to locate the transmitter in the network and thus perform jamming. To solve this problem, the covert communications technique is proposed as a more demanding PLS technique. In order not to be detected by malicious nodes in the act of transmitting data, transmitters can hide the transmitting activities by designing a suitable power allocation scheme or by using friendly jammers. A comparison of the aforementioned wireless communication security techniques is presented in Fig.~\ref{Visio-Compare}.

However, current research on PLS, covert communications, and encryption is mainly carried out in the conventional IoT without considering the new features of SIoT. Moreover, the study of the security of semantic communications techniques is still in its infancy~\cite{luo2022encrypted,hu2022robust,yang2022semantic}. The application of classical wireless communication security techniques in SIoT has not been clearly discussed. Specifically, the following questions have not been answered:
\begin{enumerate}
    \item[{\bf{Q1)}}] How do wireless communication security techniques differ in the SIoT compared to those in the conventional IoT?
    \item[{\bf{Q2)}}] What are the security performance indicators in SIoT?
    \item[{\bf{Q3)}}] What new security issues does semantic communication technology bring while improving the efficiency of the network?
\end{enumerate}
As such, we revisit classical communication security techniques from the perspective of semantic networks, and discuss the novel attack and defense methods brought about by the semantic communications techniques. Our contributions are summarized as follows:
\begin{enumerate}
    \item We revisit three security techniques, i.e., PLS, covert communications, and encryption. For each technique, we discuss its new features in SIoT (For {\bf{Q1}}).
    \item To quantify the new characteristics brought by the SIoT to PLS and covert communications, we propose two new performance indicators, i.e., semantic secrecy outage probability (semantic SOP) and detection failure probability (DFP), respectively (For {\bf{Q2}}).
    \item We discuss the semantic attack schemes caused by semantic communications technique, which can be divided into targeted and untargeted semantic attacks (For {\bf{Q3}}). Furthermore, we propose training-based and training-free defense schemes.
\end{enumerate}

\section{Revisiting Conventional Security Techniques}
In this section, we revisit wireless communication security techniques, including PLS, covert communications, and encryption techniques. We present the definitions, features, and common security performance indicators. We then discuss the differences between these techniques when applied in the SIoT and in the conventional IoT, and propose novel performance indicators.
\subsection{Physical Layer Security}
\subsubsection{Definition}
The PLS is built upon the information theory, which aims to protect wireless communications against eavesdropping by exploring and utilizing the inherent features of the physical medium~\cite{chorti2022context}. Unlike the encrypt method, PLS is independent of device computing capability, which not only enables it to achieve effective security but also gives it a natural advantage in saving resources. Moreover, such a technique is able to adjust transmission strategies according to the physical layer characteristics to adapt to the wireless channel changes. According to the working principles of PLS methods, the security performance indicators mainly include secrecy outage probability (SOP), the probability of non-zero secrecy capacity (PNZ), and the average secrecy capacity (ASC)~\cite{mukherjee2014principles}. 

\begin{itemize}
	\item {\bf{Average Secrecy Capacity:}} The secrecy capacity can be obtained by calculating the difference between the main channel capacity and wiretap channel capacity. For a given constraint of perfect secrecy, the average secrecy capacity provides a criterion for capacity limit from a mathematical point of view. For instance, if the legitimate user is able to obtain the perfect channel state information (CSI) of the eavesdropper’s channel, then the coding scheme can be flexibly adjusted to adapt to different fading coefficients. In principle, therefore, one can realize any average secure communication rate, which is below the average secrecy capacity of the channel.
\item {\bf{Secrecy Outage Probability:}} The SOP is defined as the probability that the instantaneous secrecy capacity falls less than the target secrecy rate~\cite{mukherjee2014principles}. The SOP first provides the conditions that the wireless channel needs to meet to support the specified secure rate. Second, it gives a security measure for cases where legitimate users have no CSI about the eavesdropper. Therefore, as long as the secrecy capacity is larger than the target secrecy rate, the eavesdropper’s channel is worse than legitimate users’ estimation and the secrecy of the network is ensured.
\item {\bf{Probability of Non-Zero Secrecy Capacity:}} If the main channel capacity is larger than that of the eavesdropper’s channel, the eavesdroppers are unable to successfully decode the transmitted information. Here, the occurrence probability of such an event is defined as the PNZ. On this basis, according to the definition, the PNZ equals the probability that the instantaneously received signal-to-noise ratio (SNR) of the legitimate user is greater than that of the eavesdropper.
\end{itemize}

\subsubsection{Physical Layer Security in SIoT Network}~\label{PLSS}
\begin{figure}[t]
\centering
\includegraphics[width = 0.48\textwidth]{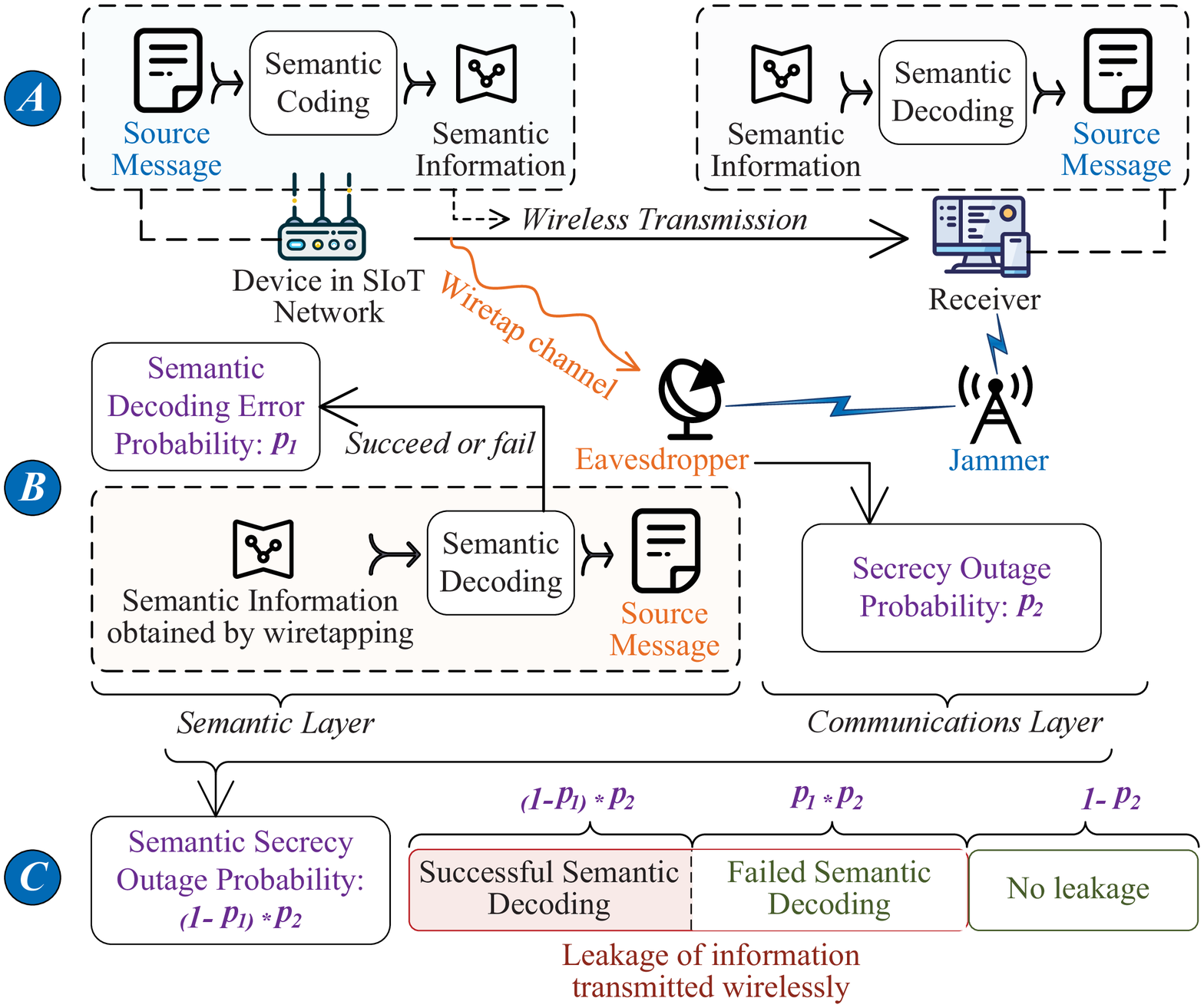}
\caption{Rethinking physical layer security technique in the semantic Internet of Things. A novel security performance indicator, i.e., semantic secrecy outage probability, is proposed.}
\label{Visio-PLS}
\end{figure}
In a conventional IoT, an eavesdropper is considered to be successful in eavesdropping if it can obtain the source message sent by the transmitter. However, in the SIoT, semantic information is transmitted in the wireless channel and decoding is required to obtain the original source message, as shown in Fig.~\ref{Visio-PLS} (Part A). Therefore, even if the eavesdropper intercepts the semantic information sent by the transmitter, it may still not obtain the required information, as shown in Fig.~\ref{Visio-PLS} (Part B). 
According to the characteristics of semantic communication, we discuss the possible scenarios in which the eavesdropper succeeds in eavesdropping on information but fails in decoding it as follows:
\begin{enumerate}
    \item[a)] {\textit{Semantic decoding failure caused by background knowledge difference between the eavesdropper and the legitimate receiver.}} In the SIoT, legitimate communication participants will share the background knowledge for semantic encoding and decoding. The eavesdropper might not have the same background knowledge to decode the intercepted semantic information. For example, the semantic information ``Mouse'' could be decoded as an animal or a computer device, in which the eavesdropper may not be able to decode this meaning correctly. Moreover, the dataset used for semantic encoder and decoder training can be also regarded as the shared background knowledge that is unavailable to the eavesdropper.  
    \item[b)] {\textit{The eavesdropper's task objectives are different from those of the legitimate receiver, resulting in the eavesdropper's inability to obtain the desired information.}} For example, the transmitter is a camera in the SIoT that captures street view photos. With the help of semantic communication technology, the camera acts as a transmitter to send these photos to a legitimate receiver for storage. The owner of the legitimate receiver is a vehicle company interested in the number and type of cars on the street, while an eavesdropper wants to steal pedestrians' information on the road from images. However, the transmitter, through the semantic encoder, transmits only the semantic information that satisfies the task of the legitimate receiver, e.g., image segmentation of vehicles. Therefore, even if the eavesdropper has perfect access to the images transmitted in the wireless environment, its task requirement is not met.
    \item[c)] {\textit{Semantic decoding failure caused by the encryption mechanism of the semantic encoding model itself.}} Due to the fact that the semantic communication technique requires only partial data to be transmitted and the decoding of semantic information relies on the receiver’s decoder design, it has also been regarded as a potential method for secure communications, the eavesdroppers are unable to decode successfully. The reason is that the eavesdropper might not have the semantic decoder that is jointly trained with the semantic encoder to recover the original source message.
\end{enumerate}
Based on the above discussion, one can conclude that the metrics reflecting the security performance of the SIoT are not only SOP at the communication level but also the {\textit{semantic decoding error probability}} (SDEP) at the semantic level. Here SDEP describes the probability that the eavesdropper can successfully decode the information that it needs from the intercepted semantic information. For example, in a semantic communications system with visual question answering (VQA) task~\cite{xie2021task}, SDEP can be the probability that the eavesdropper uses the eavesdropped semantic information model to obtain the correct answers to its own questions.
To measure the security performance in a comprehensive manner, we propose a new security performance metric for physical later security technique in the SIoT, which is named {\textit{semantic SOP}}.
\begin{rem}
    As shown in Fig.~\ref{Visio-PLS} (Part C), semantic SOP describes the probability that the eavesdropper in the SIoT successfully eavesdrops on the semantic information sent by the transmitter and successfully performs the semantic decoding. Therefore, semantic SOP is defined as the product of {\textit{SOP}} and {\textit{one minus SDEP}}.
\end{rem}
Note that when the task or interest of the eavesdropper is unknown to the system designer, semantic SOP cannot be calculated accurately by the SIoT designer. However, our proposed semantic SOP can still be used for theoretical upper bound performance analysis, to verify the robustness of the proposed system design. Another feasible research direction is to use statistical methods to estimate the semantic information that eavesdroppers may wish to obtain. Thus, the system designer can estimate SDEP to aid in secure SIoT design, with the help of our proposed semantic SOP metric.


\begin{figure}[t]
\centering
\includegraphics[width = 0.48\textwidth]{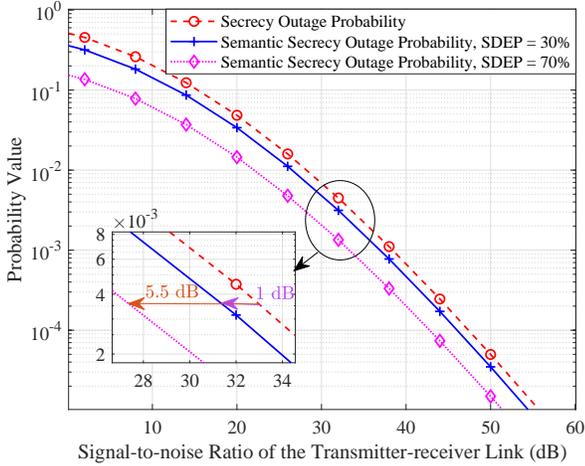}
\caption{The SOP and the semantic SOP versus the signal-to-noise ratio of the transmitter-receiver link with different values of SDEP. The distance between the transmitter and the receiver is $15$ ${\rm{m}}$, the distance between the transmitter and the eavesdropper is $18$ ${\rm{m}}$, the signal-to-noise ratio of the transmitter-eavesdropper link is $0$ ${\rm{dB}}$, the target secrecy rate threshold is $1$, the number of antennas for the transmitter, receiver, and eavesdropper are all $3$, and the path loss exponent is $2$.} 
\label{SSOP}
\end{figure}
Figure~\ref{SSOP} shows the SOP and the semantic SOP versus the signal-to-noise ratio of the transmitter-receiver link with different values of SDEP. We can observe that the probability that the eavesdropper fails to decode the semantic information reduces the semantic SOP. Specifically, when an eavesdropper in the semantic IoT has a $30\%$ probability of not being able to decode successfully semantic information, i.e., SDEP = $30\%$, the transmitter can use an SNR roughly 1 ${\rm{dB}}$ lower to achieve the same SSOP as the SOP in a conventional IoT. Moreover, if a better pair of semantic encoder and decoder can be designed to increase further the error probability of the eavesdropper in decoding the semantic information, e.g., SDEP = $70\%$, we can observe that the transmitter can use an SNR 5.5 ${\rm{dB}}$ lower to achieve the same SSOP as the SOP in a conventional IoT.

An important inspiration for secure SIoT research is to perform cross-layer co-design. Although semantic communications can enhance the system security and reduce the transmit power required to achieve a certain SSOP, the encoding and decoding of semantic information consume computational resources of the network. Therefore, the tradeoff between SIoT performance and security should be considered. By jointly designing the transmit and jamming power allocation schemes in the physical layer and the semantic encoding/decoding scheme in the semantic layer, a more secure and more efficient IoT can be achieved.

\subsection{Covert Communications}
\subsubsection{Definition}
So far, the PLS has been applied at large to boost wireless transmission security. Despite its effectiveness, PLS still has certain limitations in other aspects. By analyzing the wireless signal, for instance, the user's location may be exposed, which poses threat to user privacy. Such problems cannot be solved by PLS techniques, triggering the proposal of covert communications. Also known as low probability of detection communications, covert communications aim to deliver information to a legitimate user without being caught by the warden, who attempt to detect such transmission~\cite{makhdoom2022comprehensive}. The covert communications can include two major aspects. The first one mainly focuses on analyzing and exploiting the uncertainty of the average power of malicious wardens. Another one is to send the signal covered by high-power signals, so as to improve covertness. It is not difficult to see that covert communication never relies on the adversary’s competence, indicating that transmission security can be perfectly guaranteed even if the attacker has a strong processing capability. According to the above discussion, the covert rate and DEP, which are detailed as follows, are used to characterize the performance of covert communications.
\begin{itemize}
\item {\bf{Detection Error Probability:}} The warden needs to make a binary choice between silent and transmitting via hypothesis testing. Therefore, the detection error probability (DEP) is defined as the likelihood of the warden making a wrong decision, which contains two cases. The first one is that the warden chooses non-null-decision (transmitting) while the null hypothesis (silent) is true, which is called {\textit{false alarm}}. Another one is that warden sides with a null hypothesis when the non-null hypothesis is true, which is known as {\textit{miss detection}}. The value of DEP is the sum of the probabilities of making the above two wrong decisions.
\item {\bf{Covert Rate:}} Besides DEP, covert rate, which describes the data transmission rate when the DEP of the warden is close to one, is also vital. The covert rate of any user can be calculated based on the well-known Shannon–Hartley theorem~\cite{rioul2014shannon}.
\end{itemize}
\subsubsection{Covert Communications in SIoT Network}
\begin{figure*}[t]
\centering
\includegraphics[width = 0.95\textwidth]{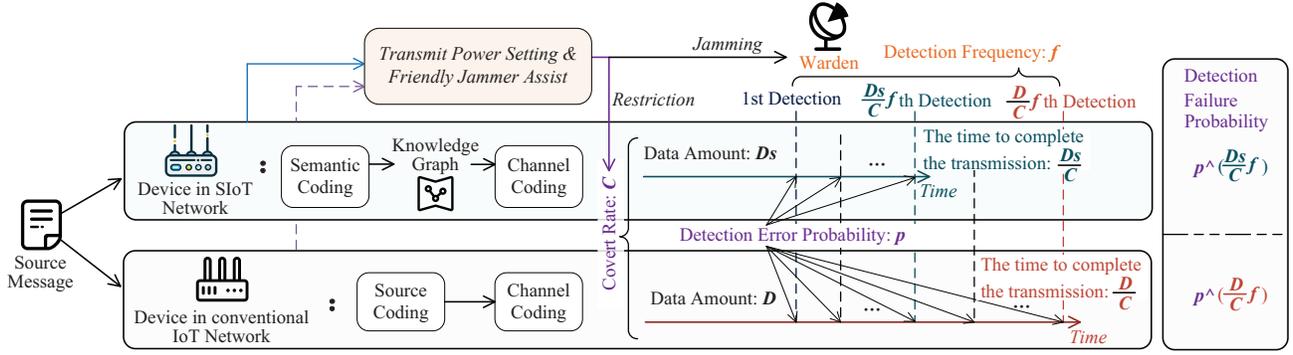}
\caption{Rethinking covert communications technique in the semantic Internet of Things. A novel security performance indicator, i.e., detection failure probability, is proposed.}
\label{Visio-Covert2}
\end{figure*}
In a covert communication system, the objective of the warden is to detect whether the transmission is taking place or not, without caring what data is being transmitted. Therefore, encoding the source message to be transmitted into semantic information will not improve the DEP of the warden. If the warden successfully detects that wireless communication is taking place, it can analyze and obtain the transmitter's information such as the location, and then apply interference to block the semantic communications.
To make the DEP converge to 1 arbitrarily, the solutions are to design a reasonable transmitting power allocation scheme and/or to use a friendly jammer or a reconfigurable intelligent surface, as in the conventional IoT. The covert rate in the semantic IoT is the same as that in the conventional IoT.

However, because the warden can perform multiple detections during the data transmission process, a failure of one detection does not mean that the warden cannot discover the transmitting activity. Although the DEP can be arbitrarily close to 1 with covert communication techniques, the DEP is typically set as 90\% to 95\% in practical communication systems~\cite{zheng2021wireless}. Even if we consider that the DEP is $99\%$, the probability that a warden who can detect the wireless environment $5$ times per second finds a transmitting activity that lasts $10$ seconds is $1 - 99\%^{50} = 39.5\%$. We can conclude that, when transmitting the same amount of data, e.g., 1 article of 1000 words, with the same warden DEP, the probability of the wireless transmission being successfully detected by the warden is lower in the SIoT than that in the conventional IoT. The reason is that in the SIoT, the transmitter can encode articles into semantic information without affecting task completion, e.g., knowledge graphs, which have fewer bits than the original article. Therefore, with the same covert rate, transmitters in the SIoT can complete information transmission faster, as shown in Fig.~\ref{Visio-Covert2}. Because there is an upper limit to the frequency of the warden's detection of the wireless environment, the shorter the transmission time is, the lower the probability that the transmitting activity will be detected.

However, there is no suitable performance indicator to describe the difference in covert communication security performance that is caused by the data amount difference. This research gap exists because all source messages are encoded in the conventional IoT and there are no differences in the amount of data. To fill this gap, we propose a new performance indicator for covert communications, i.e., {\textit{detection failure probability}} (DFP), as follows:
\begin{rem}
DFP describes the probability that no transmitting activity is detected by the warden's multiple detections during the transmission time of the data. Therefore, as shown in Fig.~\ref{Visio-Covert2}, DFP can be defined as a power function of DEP, where the power is the number of detections. Considering that the warden performs $f$ detections per unit time due to energy constraints, the number of detections can be calculated as the data amount divided by covert rate and then multiplied by $f$.
\end{rem}

\begin{figure}[t]
\centering
\includegraphics[width = 0.48\textwidth]{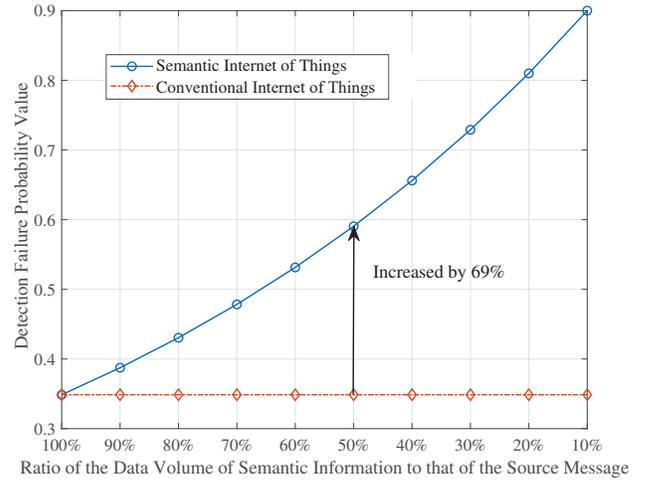}
\caption{The detection failure probability versus the ratio of the data volume of semantic information to that of the source message, in the SIoT and conventional IoT, respectively. The DEP is $95\%$, the covert rate is $20$ ${\rm bit/s/Hz}$, the bandwidth is $5$ ${\rm MHz}$, and the warden detects the wireless environment at a frequency of twice per second.}
\label{Covert}
\end{figure}
Figure~\ref{Covert} illustrates the detection failure probability versus the ratio of the data volume of semantic information to that of the source message. We can observe that although the warden's DEP is the same in both the SIoT and conventional IoT, i.e., 90\% as is set in much of the literature~\cite{zheng2021wireless}, a great gap exists in the achievable DFP. For example, if the semantic encoder can reduce the number of bits of the source message by half to obtain the semantic information, the DFP can be improved by $69\%$. 

An interesting insight is that there is a tradeoff between the semantic encoder computing resource and the physical layer transmit power resource. However, unlike the tradeoff we discussed in Section~\ref{PLSS}, for the semantic encoder in the covert communication-aided SIoT, what matters is how much the encoder reduces the number of semantic information bits. For the semantic encoder in the PLS-aided SIoT, what matters is how low the probability is that the semantic information is decoded successfully by an illegal eavesdropper.

\subsection{Encryption}
\subsubsection{Definition}
Encryption is one of the most classical techniques to ensure secure transmission, which operates in the upper layers of the communication system. 
For encryption techniques, there are comprehensive metrics for performance evaluation, including but not limited to crack time, the throughput of encryption/decryption, and power consumption. The crack time and the computational resources to be used by the eavesdropper are positively related to the key size.

\subsubsection{Encryption in SIoT Network}
In the SIoT, encryption techniques can be seen as a ``second layer'' of protection for the transmitted data. The reason is that the input of the encryption algorithm can be semantic information obtained through semantic coding, and semantic information itself has encryption properties. Even if the eavesdropper succeeds in breaking the encryption, it may not succeed in decoding the semantic information to obtain the source message, as we discuss in Section~\ref{PLSS}.

A general design is to integrate cryptography as an option with semantic communication systems \cite{luo2022encrypted}. In SIoT, if the transmitter and receiver want to hide the information from a potential eavesdropper, the goal is to minimize the error between the transmitter and receiver while maximizing the error between the transmitter and eavesdropper. Following this idea, an encrypted semantic communication system is designed in \cite{luo2022encrypted}. However, the authors in \cite{luo2022encrypted} only considered symmetric encryption, and related work is still in the vacant stage. More cryptography-assisted semantic communication systems can be designed to further improve the security of SIoT.

\section{New Security Issues}
In this section, we discuss the new security issues arising from the introduction of semantic communication techniques in the IoT.

\subsection{Semantic Attack in SIoT Network}
Unlike the bit streams transmitted in conventional IoT, the semantic information in SIoT is largely task-related and dependent on the design of the semantic encoder and decoder. However, a variety of error-correcting coding methods have been designed to correct bit errors, but methods that can reduce semantic noise have rarely been investigated. The semantic noise can have a small or large impact on system performance. For example, because of the small deviation of the text semantic vectors, the receiver decodes ``bike'' into ``bicycle'' when recovering the text message. The receiver's judgment will not be affected. However, the disturbance of some semantic information may seriously affect the communication system. For example, if the images are incorrectly semantic encoded and uploaded to a dataset, the quality of the artificial intelligence model trained by the dataset may be affected.

The mismatch between the original source message and the obtained semantic information by semantic encoding is called semantic noise~\cite{hu2022robust}, which is considered as a special type of noise present in semantic communication systems. In the SIoT, some semantic noise is naturally present, e.g., different users have different interpretations of the same word, and require better semantic encoding and decoding design to overcome. However, some semantic noise is generated by attackers with the aim of disrupting the semantic communication system.
For source messages in text form, synonym substitution or reversing the order of certain letters may cause the deep learning-based semantic model to misinterpret the semantics of the sentence. 
For source messages in image form, only by changing some pixels in an image, the semantic information extracted by a well-pre-trained semantic encoder can be completely inconsistent with the real content of the image~\cite{szegedy2014intriguing}. Regardless of the modality of the source message, the goal of the semantic attack can vary and corresponds to different loss function optimization.

\begin{figure*}[t]
\centering
\includegraphics[width = 0.95\textwidth]{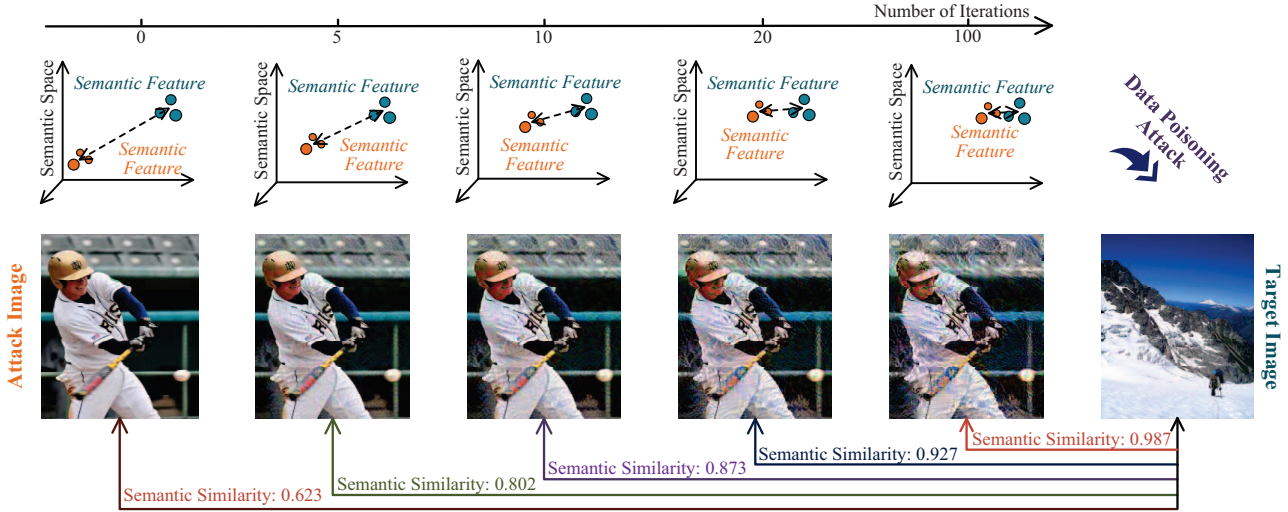}
\caption{The targeted semantic attack approach. The experimental platform for running the attack algorithm is built on a generic Ubuntu 20.04 system with an AMD Ryzen Threadripper PRO 3975WX 32-Cores CPU and an NVIDIA RTX A5000 GPU.}
\label{attackshow}
\end{figure*}
\begin{itemize}
\item {\bf{Targeted Semantic Attack:}} The goal is to generate semantic tampered source messages with a given target semantic information. Here, the target semantic information is the semantic information that the receiver in the SIoT wants to receive. For example, a digital twin service provider wants to collect some images with {\textit{snowy mountain}} as semantic information to build a virtual object. The attacker can change some pixels in an irrelevant image to make the semantic information of the irrelevant image very close to {\textit{snowy mountain}}. Therefore, the loss function could be the cosine similarity of the semantic vectors of the irrelevant pictures and the snowy mountain pictures. As shown in Fig.~\ref{attackshow}, as the number of iterations increases, the semantic similarity of the two images gradually increases. If the digital twin service provider cannot correctly detect the semantic tampered images, its database will be contaminated. Therefore, this type of attack can be called {\textit{semantic data poisoning attack}}. Such a semantic tampering approach can also be used for man-in-the-middle attacks. A malicious intermediate node capable of intercepting the wireless communication channel can replace the images to be transmitted, without affecting the semantic information.
    \item {\bf{Untargeted Semantic Attack:}} Similar to the approach of the target semantic attack but with a different objective function, the aim of the untargeted semantic attack is to minimize the similarity between the semantic information of the tampered source message and its true semantic information. In this case, the devices in the SIoT are unable to perform properly the semantic encoding of the maliciously tampered source message. For example, the semantic feature of the attack image in Fig.~\ref{attackshow} is not iterating to be closer to the ``snowy mountain'', but as far from the ``baseball player'' as possible.
\end{itemize}

\subsection{Defense Methods}
\subsubsection{Training-based Defense}
By considering that a large part of the semantic communication encoders and decoders are functioned by deep learning methods, a feasible solution to reduce semantic noise is to improve the robustness of semantic models during the training process. Specifically, there have been some training-based methods, e.g., defensive distillation~\cite{papernot2016distillation}, weight perturbation~\cite{wu2020adversarial}, and adversarial training~\cite{bai2021improving}.

Although there have been several approaches to improve model robustness using adversarial samples in the fields of natural language and image processing~\cite{bai2021improving}, semantic noise-resistant models in wireless communication have not been sufficiently studied. Fortunately, researchers can draw inspiration from existing adversarial training methods and consider the impact of wireless channels and transmission overhead in the SIoT. Most recently, to reduce the impact of semantic noise on the system, a masked vector quantized-variational autoencoder (VQ-VAE) is developed as the architecture of the robust semantic communication system~\cite{hu2022robust}. To improve the system robustness, a feature importance module is proposed to suppress noise-related and task-independent features. It is shown that the proposed masked VQ-VAE requires $0.36\%$ transmitted symbols of the conventional ``joint photographic experts group (JPEG) + low-density parity-check coding'' method~\cite{hu2022robust}, while effectively
improving the system robustness by reducing the impacts of semantic noise.

\subsubsection{Training-free Defense}
Research on training-free defense methods remains to be developed. There may be different defense methods for data with different modalities. Taking image data as an example, a possible defense solution is to use the visual invariance of the semantic tampered images for correct semantic extraction. As shown in Fig.~\ref{attackshow}, although the semantic similarity between the two images is high for a semantic encoder, the human eye can easily see the difference between them. The reason is that only some pixels in the attack image have been adjusted. Inspired by this, we try to blur both images by using Gaussian method, and find that the semantic similarity between them can be reduced from $0.987$ to $0.78$, without retraining the semantic model. More pre-processing solutions could be investigated to defend against this kind of semantic attack as a future research direction.

\section{Future Direction}
\subsection{Explainable AI-aided Semantic Communications}
The past few years have witnessed the rapid development of machine learning (ML) technologies, especially deep learning, which has shown significant advantages in a variety of applications. Most of them, however, are often unable to explain their decisions and actions during the operation process, triggering the research on explainable artificial intelligence (XAI). The XAI aims to provide users with detailed explanations of how the decision is made or the result is obtained. For semantic communications, which rely on ML, the XAI can make the training of transceiver pairs change from black box to white box, driving the training process clearer and easier to understand. Such improvement not only allows semantic communication system designers to identify and fix potential vulnerabilities or threats but also helps users understand and trust these semantic communications better. Therefore, studying explainable AI-aided semantic communications is an indispensable step to improve its security.
\subsection{Blockchain-aided Semantic Internet of Things Network}
The blockchain is a chain of blocks that store all committed transactions in a decentralized and distributed network. Unlike the conventional ways, the blockchain realizes the peer-to-peer digital assets transfer without any intermediaries, and the features of decentralization, immutability, audit-ability and transparency drive the transactions' security. Considering the above advantages, the stored and shared transaction information in the blockchain can be replaced with semantic information that needs to be transmitted. In this way, not only the storage consumption of the blockchain is reduced, but also, the decentralized blockchain verification mechanism would further improve the security of semantic content. Therefore, how to better integrate blockchain and the SIoT is also worthy of further study.

\subsection{Secure Semantic Communications for Metaverse}
As a novel type of Internet application and social form, the Metaverse has received more and more attention in recent years. To provide an ideal immersive experience for users, the data that describes the user and physical world must be transmitted to the virtual world quickly and accurately, during this process effective tracking and accurate prediction are critical. Thereby, semantic communication is one of the best-suited techniques for this task. Considering the characteristics of data and the requirements of applications, it is vital to study the security of semantic communication under the framework of the Metaverse. This not only affects the user experience but is also directly related to the user's security and privacy in the physical world.

\section{Conclusion}\label{Con}
Considering the new features of the SIoT compared to the conventional IoT, we rethought three wireless communication security techniques, i.e, PLS, covert communications, and encryption. We discussed the characteristics of each technology when applied in SIoT. Specifically, because of the encryption characteristics of semantic information itself, the security enhancement that PLS can bring to the network is further improved. In addition, since semantic information that has fewer bits can be transmitted faster than the original source message, the difficulty of achieving covert communications is reduced. To show these two new characteristics, we proposed two new indicators, i.e., semantic SOP and DFP. Finally, we discussed new attacks and defense schemes that have emerged at the semantic level. We foresee that the combination of semantic communications and classical security techniques will revolutionize the architecture of communication networks, bringing new inspiration to the two most important dimensions of performance and security.

\bibliographystyle{IEEEtran}
\bibliography{Ref}
\end{document}